\documentclass[conference]{IEEEtran}
\IEEEoverridecommandlockouts
\usepackage{cite}
\usepackage{amsmath,amssymb,amsfonts}
\usepackage{algorithm}
\usepackage{algorithmic}
\usepackage{graphicx}
\usepackage{textcomp}
\usepackage{xcolor}
\usepackage{url}
\usepackage{enumitem}
\def\BibTeX{{\rm B\kern-.05em{\sc i\kern-.025em b}\kern-.08em
    T\kern-.1667em\lower.7ex\hbox{E}\kern-.125emX}}
\graphicspath{{fig/}}

\newcommand{\msg}[1]{\textsc{#1}}
\newcommand{\Neager}{\mathcal{N}^{\mathrm{E}}}
\newcommand{\Nlazy}{\mathcal{N}^{\mathrm{L}}}

\begin{document}

\title{Off-Tree Forwarding for Accelerating Broadcasts from Arbitrary Sources in Plumtree
\thanks{This work was supported by JSPS KAKENHI Grant No. 24K14935 and No. 24H00691

\textbf{This work has been submitted to the IEEE for possible publication. Copyright may be transferred without notice, after which this version may no longer be accessible.}
}
}

\author{\IEEEauthorblockN{Ryohei Banno}
\IEEEauthorblockA{\textit{Graduate School of Social Data Science, Hitotsubashi University}, Tokyo, Japan \\
banno@computer.org}
}

\maketitle

\begin{abstract}
Overlay broadcast is fundamental to large-scale distributed systems, enabling messages to be disseminated to all participants.
Plumtree is a well-established protocol that embeds a spanning tree in a gossip overlay: payloads are disseminated along tree links, while non-tree links support recovery and optimization.
However, because the tree is constructed from the first broadcast, it is optimized for the initial source and can introduce long detours for broadcasts originating at other nodes, particularly leaf nodes.
To address this limitation, we propose probabilistic exuding, an off-tree forwarding mechanism in which nodes near the source additionally forward payloads over their non-tree links with a specified probability.
Exuded messages use a dedicated message type and do not participate in tree construction or repair, leaving the shared tree unchanged.
Simulations on a 1000-node random regular overlay show that, for randomly selected sources and a representative parameter setting, the proposed method reduces the 95th-percentile delivery latency by approximately 40\% relative to Plumtree.
This closes approximately 92\% of the latency gap between Plumtree and flooding while sending only about 58\% as many payload messages as flooding.
In contrast, Plumtree's own link optimization and shorter repair timeouts, which also exploit non-tree links, rewire the shared tree and increase latency as broadcast sources change; the latency of exuding is nearly unaffected.
\end{abstract}


\section{Introduction}

Disseminating information to large numbers of nodes is a fundamental problem in distributed systems, dating back to epidemic algorithms for replicated databases~\cite{demers1987epidemic}.
It remains important in blockchains~\cite{vyzovitis2020gossipsub}, UAV-assisted edge computing~\cite{ren2024intelligent}, and cloud data replication~\cite{wooders2024cloudcast}.
Because IP multicast has not been widely deployed, application-layer multicast has been studied extensively~\cite{jannotti2000overcast,chu2002case,banerjee2002scalable,castro2002scribe,hosseini2007survey}.
Tree-based schemes generally deliver each message to each node once but are vulnerable to failures, whereas gossip-based schemes forward messages redundantly, improving reliability at the cost of additional traffic~\cite{birman1999bimodal,eugster2003lightweight,rodrigues2003adaptive}.

Hybrid protocols~\cite{tang2005gocast,leitao2007epidemic} combine tree-based efficiency with gossip-based resilience.
Plumtree~\cite{leitao2007epidemic}, a well-established example of such a hybrid protocol, forms a spanning tree from the first broadcast, forwarding payloads along tree links and message identifiers over non-tree links.
Although efficient for the initial source, the shared tree can impose long detours on broadcasts from other sources.
Its optional hop-count-based tree optimization may favor one source at another's expense (Section~\ref{sec:plumtree}).
Our simulations show that this optimization and shorter repair timeouts can increase latency when sources change.

Inspired by PRM~\cite{banerjee2006resilient} and FPCast~\cite{li2009fast}, we propose probabilistic exuding, which forwards payloads over non-tree links near each broadcast source with a specified probability.
Dedicated message types separate exuded traffic from tree maintenance, preserving the shared tree.

Our contributions are:
\begin{enumerate}[label=(\roman*)]
  \item A protocol that separates off-tree payload forwarding from Plumtree's tree maintenance, controlled by a forwarding probability $p$ and a hop threshold $\theta_h$.
  \item Simulation evidence of reduced latency with bounded payload traffic across different broadcast sources, overlay sizes and degrees, and bandwidth conditions.
  \item Evidence that concentrating exuding near the source with $p=1$ yields a better latency--traffic trade-off than forwarding with lower probability over more hops.
\end{enumerate}

\section{Related Work}
\label{sec:related}

Overlay broadcast protocols balance traffic and resilience. Gossip-based schemes use redundant forwarding~\cite{demers1987epidemic,birman1999bimodal,eugster2003lightweight,rodrigues2003adaptive}.
GossipSub~\cite{vyzovitis2020gossipsub} has also been studied in the context of the XRP Ledger~\cite{scheidt2023pubsub,scheidt2024nine}.
Adaptive and cost-aware approaches address dissemination efficiency~\cite{ren2024intelligent,wooders2024cloudcast}.
Tree-based designs such as Overcast~\cite{jannotti2000overcast}, Narada~\cite{chu2002case}, NICE~\cite{banerjee2002scalable}, and Scribe~\cite{castro2002scribe} reduce redundant payload traffic but are vulnerable to failures~\cite{hosseini2007survey}.
Hybrid protocols such as GoCast~\cite{tang2005gocast}, Plumtree~\cite{leitao2007epidemic}, Thicket~\cite{thicket2010}, and BRISA~\cite{brisa2012} combine tree-based efficiency with gossip-based resilience.
HyParView~\cite{leitao2007hyparview} maintains overlay connectivity, while X-BOT~\cite{xbot2009} optimizes overlay links.
Plumtree, used in OpenRiak\footnote{\url{https://github.com/OpenRiak/} (accessed Oct. 15, 2026)} and Bondy\footnote{\url{https://github.com/bondy-io/} (accessed Oct. 15, 2026)}, supports both per-source trees and a single shared tree for multiple broadcast sources.
This paper focuses on the latter configuration, which avoids per-source tree maintenance at the cost of potentially longer delivery paths.

PRM~\cite{banerjee2006resilient} and FPCast~\cite{li2009fast} also employ probabilistic forwarding.
PRM forwards to randomly selected overlay nodes for resilience, whereas FPCast uses DHT-based rendezvous for multi-source multicast.
We focus on broadcast over an existing unstructured overlay using only neighbor links.
Direct comparisons would require adapting their forwarding models or underlying structures.
We therefore compare exuding with Plumtree and flooding on the same overlay and initial tree to isolate the effect of off-tree forwarding (Section~\ref{sec:evaluation}).

A shared tree avoids per-source maintenance but can lengthen delivery paths from sources other than the one that formed it~\cite{leitao2007epidemic}.
We exploit existing non-tree links as per-broadcast shortcuts without altering the shared tree.
Section~\ref{sec:discussion} discusses per-source trees.

\section{Plumtree}
\label{sec:plumtree}

In Plumtree~\cite{leitao2007epidemic}, each node $v$ partitions its neighbors into \emph{eager push peers} $\Neager_v$ and \emph{lazy push peers} $\Nlazy_v$; initially all are eager.
On first receiving a message $m$ with the message type \msg{Gossip}, $v$ delivers it, forwards the payload as a \msg{Gossip} message to $\Neager_v$ except the sender, and sends an \msg{IHave} message with only the identifier to $\Nlazy_v$.
Messages carry a hop counter; $h_m(v)$ is the hop count of the path by which $v$ first received $m$.
On a duplicate \msg{Gossip}, $v$ moves the sender to $\Nlazy_v$ and replies with \msg{Prune}, and the sender does the same.
The first broadcast thus prunes all redundant eager links, leaving a spanning tree shaped by the paths of that message.
We also refer to eager links as tree links and to lazy links as non-tree links.
On an \msg{IHave} for a missing message, $v$ starts a timer $T_1$; on expiry, it sends \msg{Graft} to the first announcer to repair the tree and retrieve the payload.
If the payload is still missing after a shorter timer $T_2$, it retries with the next announcer, if any.
A missing \msg{Gossip} thus also reveals a broken tree.

Plumtree offers an optimization intended to exploit better paths that appear as the overlay evolves.
If an \msg{IHave} for $m$ arrives with a hop count that is smaller than that of the \msg{Gossip} by at least a threshold, the node grafts the lazy link and prunes the eager one.
The authors also apply this optimization to a tree shared by multiple senders~\cite{leitao2007epidemic}.
With multiple sources, a swap that helps one source may hurt another.
For a tree $A$--$B$--$C$--$D$ with a lazy link $D$--$A$ and sources $B$ and $D$, making $D$--$A$ eager speeds up $D$'s messages to $A$ but turns $A$--$B$ lazy and lengthens $B$'s path, so the tree may oscillate.
The authors therefore recommend a threshold close to the diameter for multiple senders, to avoid constant link changes~\cite{leitao2007epidemic}.

\section{Probabilistic Exuding}
\label{sec:proposal}

\subsection{Idea}

A node $v$ that sends a broadcast $m$, or first receives it from a neighbor $u$, additionally forwards $m$ to each $w \in \Nlazy_v \setminus \{u\}$ independently with probability $p$ if $h_m(v) \le \theta_h$, where $h_m(v)=0$ at the source.
We call this \emph{exuding}.
With $\theta_h=0$ only the source exudes, and with $p=1$ and a large $\theta_h$ the payload is disseminated as in flooding.
In the variant evaluated in Section~\ref{sec:decay}, the forwarding probability decreases with the hop count, $p_h = p\,\gamma^h$ with $\gamma \in (0,1]$.

\begin{figure}[!t]
\centering
\includegraphics[width=0.7\linewidth]{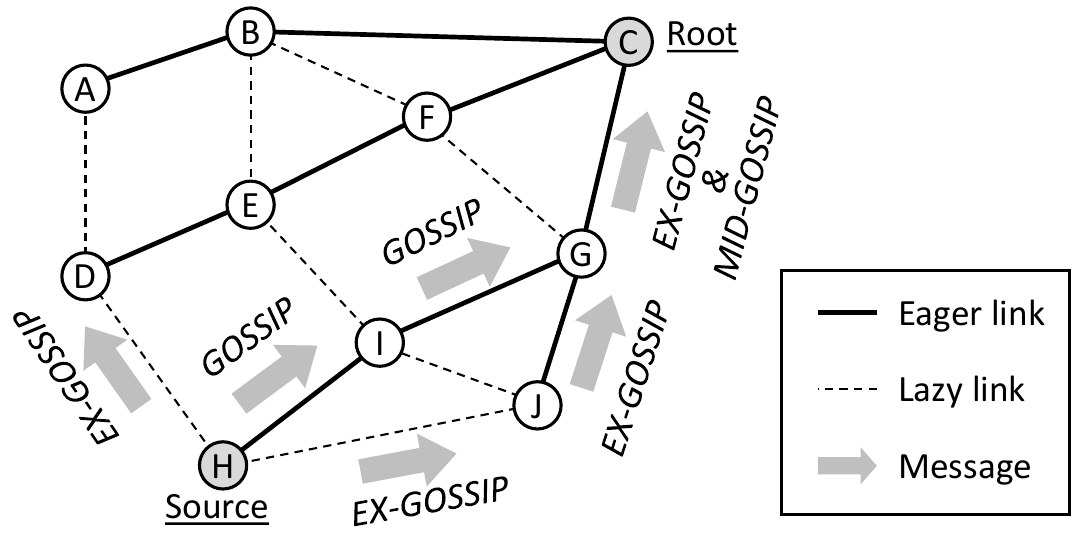}
\caption{Example of exuding.}
\label{fig:overview}
\end{figure}

Fig.~\ref{fig:overview} shows an example.
Node $C$ issued the first broadcast, which formed the spanning tree.
When node $H$ broadcasts, it additionally sends \msg{Ex-Gossip} messages, each with probability $p$, to its lazy push peers $D$ and $J$; an \msg{Ex-Gossip} to $D$ arrives in one hop, whereas Plumtree needs six hops along the tree ($H$--$I$--$G$--$C$--$F$--$E$--$D$).

\subsection{Keeping the tree intact}

Exuded copies cannot simply be sent as \msg{Gossip}: the resulting duplicates would trigger \msg{Prune} and reshape the tree around every source.
Nor can a node that has already received the payload off-tree simply stop forwarding along the tree: downstream nodes would then never receive the tree copy, so redundant links would no longer be pruned, a real break would go unnoticed, and timeouts at nodes without the payload would trigger spurious \msg{Graft}s on an intact tree.
We therefore introduce three message types:
\msg{Ex-Gossip} carries an exuded payload and is never used for pruning or grafting;
\msg{Mid-Gossip} carries only the identifier along the tree, replacing \msg{Gossip} at nodes that have already received the payload via \msg{Ex-Gossip};
and \msg{Gossip-Req} asks a tree neighbor for the payload when a \msg{Mid-Gossip} arrives at a node that lacks it.

Each node keeps the identifier sets $M^{G}_v$, $M^{EX}_v$, and $M^{MID}_v$ received by \msg{Gossip}, \msg{Ex-Gossip}, and \msg{Mid-Gossip}, respectively.
Algorithm~\ref{alg:exuding} shows the handlers that differ from Plumtree.
When a node first receives a payload via \msg{Ex-Gossip}, it delivers the payload, forwards it to eager peers as \msg{Ex-Gossip}, and may exude it further.
When the \msg{Gossip} later arrives along the tree, $v$ acts as in Plumtree, i.e., it stops timers, sends \msg{IHave}, and keeps the sender eager, but forwards only a \msg{Mid-Gossip}.
In Fig.~\ref{fig:overview}, if $G$ receives the \msg{Ex-Gossip} from $J$ before the \msg{Gossip} from $I$, it first forwards the \msg{Ex-Gossip} to $C$ and later a \msg{Mid-Gossip}, through which $C$ maintains the tree as in Plumtree.
A \msg{Mid-Gossip} is treated like a \msg{Gossip} message without a payload: duplicates trigger \msg{Prune}, while the first reception causes the identifier to be forwarded down the tree.
If it arrives at a node that has not yet received the payload, e.g., because an \msg{Ex-Gossip} was lost, the node sends \msg{Gossip-Req}, to which an eager sender replies with \msg{Gossip}; if this is lost too, the usual repair applies.
On receiving an \msg{IHave}, a node treats the announced message as missing and starts the timer unless the identifier is in $M^{G}_v \cup M^{MID}_v$, so that a broken tree is still detected even when the payload arrived by \msg{Ex-Gossip}.
A \msg{Graft} indicates whether the payload is needed, and the reply is a \msg{Mid-Gossip} if the payload was already received by \msg{Ex-Gossip}.
In short, a node obtains each message's payload by \msg{Gossip} or \msg{Ex-Gossip} and its tree copy by \msg{Gossip} or \msg{Mid-Gossip}, and only the tree copy drives pruning and repair.

\begin{algorithm}[!t]
\caption{Main handlers that differ from Plumtree (changes to \msg{IHave}, timer, and \msg{Graft} handling are described in the text).}
\label{alg:exuding}
\footnotesize
\begin{algorithmic}[1]
\STATE \textbf{procedure} exude$(m, h, u)$
\STATE \quad \textbf{if} $h \le \theta_h$ \textbf{then}
\STATE \quad\quad \textbf{for each} $w \in \Nlazy \setminus \{u\}$ \textbf{with} $\mathit{random}() \le p$ \textbf{do}
\STATE \quad\quad\quad send(\msg{Ex-Gossip}, $m$, $h$) to $w$
\vspace{0.5em}
\STATE \textbf{upon} broadcast$(m)$:
\STATE \quad send(\msg{Gossip}, $m$, $0$) to $\Neager$; exude$(m, 0, \mathit{self})$
\STATE \quad send \msg{IHave} to $\Nlazy$; deliver $m$; $M^{G} \gets M^{G} \cup \{m\}$
\vspace{0.5em}
\STATE \textbf{upon} receive(\msg{Gossip}, $m$, $h$) from $u$:
\STATE \quad \textbf{if} $m \in M^{G} \cup M^{MID}$ \textbf{then} move $u$ to $\Nlazy$; send \msg{Prune} to $u$
\STATE \quad \textbf{else}
\STATE \quad\quad $M^{G} \gets M^{G} \cup \{m\}$; cancel timer of $m$
\STATE \quad\quad \textbf{if} $m \in M^{EX}$ \textbf{then} send(\msg{Mid-Gossip}, $m$, $h{+}1$) to $\Neager \setminus \{u\}$
\STATE \quad\quad \textbf{else} deliver $m$; send(\msg{Gossip}, $m$, $h{+}1$) to $\Neager \setminus \{u\}$;
\STATE \quad\quad\quad exude$(m, h{+}1, u)$
\STATE \quad\quad send \msg{IHave} to $\Nlazy \setminus \{u\}$; move $u$ to $\Neager$
\vspace{0.5em}
\STATE \textbf{upon} receive(\msg{Ex-Gossip}, $m$, $h$) from $u$:
\STATE \quad \textbf{if} $m \notin M^{G} \cup M^{EX}$ \textbf{then}
\STATE \quad\quad deliver $m$; $M^{EX} \gets M^{EX} \cup \{m\}$
\STATE \quad\quad send(\msg{Ex-Gossip}, $m$, $h{+}1$) to $\Neager \setminus \{u\}$; exude$(m, h{+}1, u)$
\vspace{0.5em}
\STATE \textbf{upon} receive(\msg{Mid-Gossip}, $m$, $h$) from $u$:
\STATE \quad \textbf{if} $m \in M^{G} \cup M^{MID}$ \textbf{then} move $u$ to $\Nlazy$; send \msg{Prune} to $u$
\STATE \quad \textbf{else if} $m \in M^{EX}$ \textbf{then}
\STATE \quad\quad $M^{MID} \gets M^{MID} \cup \{m\}$; cancel timer of $m$
\STATE \quad\quad send(\msg{Mid-Gossip}, $m$, $h{+}1$) to $\Neager \setminus \{u\}$
\STATE \quad\quad send \msg{IHave} to $\Nlazy \setminus \{u\}$; move $u$ to $\Neager$
\STATE \quad \textbf{else} send(\msg{Gossip-Req}, $m$, $h$) to $u$
\vspace{0.5em}
\STATE \textbf{upon} receive(\msg{Gossip-Req}, $m$, $h$) from $u$:
\STATE \quad \textbf{if} $u \in \Neager$ \textbf{then} send(\msg{Gossip}, $m$, $h$) to $u$
\end{algorithmic}
\end{algorithm}

\subsection{Overhead}
Exuding adds at most one payload transmission per selected lazy link.
Nodes first reached by \msg{Ex-Gossip} relay it over eager links and later forward \msg{Mid-Gossip} instead of \msg{Gossip}.
The identifiers sent along tree paths, together with the usual \msg{IHave} messages, remain necessary for pruning and failure detection.
The protocol requires only two additional message-ID sets and does not modify the eager/lazy peer sets.

\subsection{Delivery-time bounds}
\label{sec:bounds}

We establish delivery-time bounds for exuding under idealized conditions.
For a broadcast from source $s$, let $t_{\mathrm{flood}}(s,v)$, $t_{\mathrm{exude}}(s,v)$, and $t_{\mathrm{tree}}(s,v)$ be the times at which node $v$ first receives the payload under flooding, exuding, and Plumtree, respectively, on the same overlay and tree $T$ of eager links.
Exuding may use any $p$, $\theta_h$, and $\gamma$.
We assume:
(C1) the overlay does not change during the broadcast;
(C2) each link has a fixed delay that does not depend on time, message type, or load, and nodes forward without delay;
(C3) no message is lost; and
(C4) $T$ is a spanning tree that is not changed during the broadcast; in particular, no repair timer expires before the message (\msg{Gossip} or \msg{Mid-Gossip}) arrives along $T$, and no other broadcast is in progress.

\textbf{Proposition.} Under (C1)--(C4), $t_{\mathrm{flood}}(s,v) \le t_{\mathrm{exude}}(s,v) \le t_{\mathrm{tree}}(s,v)$ for every $v$.
If, in addition, (C5) $T$ was formed by the first broadcast, issued by the root $r$ with all links eager under (C1)--(C3), and has not changed since, then $t_{\mathrm{flood}}(r,v) = t_{\mathrm{tree}}(r,v)$.

\textit{Proof sketch.}
Left inequality: flooding attains the shortest-path delay, which no copy can beat.
Right inequality: under (C2)--(C4), $t_{\mathrm{tree}}(s,v)$ is the delay of the tree path $s=u_0, \ldots, u_k=v$, and on first receiving the payload every node forwards it to all eager peers but its sender, so by induction $u_i$ receives it no later than it would along this path.
Equality: a link becomes lazy at a node only through a duplicate it receives or a \msg{Prune} answering one it sent, both of which occur after its own first reception.
Hence the first broadcast is a flood without \msg{IHave} or \msg{Ex-Gossip}, and the links of first arrival, which remain eager, form a shortest-delay tree from $r$. \hfill$\square$

The relations need not hold when delays vary, the tree changes, or exuding adds queueing delays (Section~\ref{sec:bandwidth}).

\section{Evaluation}
\label{sec:evaluation}

\subsection{Setup}

\subsubsection{Common settings}
We implemented an event-driven simulator in Python using SimPy.
Unless stated otherwise, the overlay is a random regular graph with 1000 nodes and degree 8, whose diameter is 5 in all runs, resembling the symmetric, fixed-size active views of HyParView~\cite{leitao2007hyparview}.
Each link has a fixed one-way delay drawn uniformly from $[10, 100]$\,ms, with no jitter or processing delay; bandwidth is unlimited except in Section~\ref{sec:bandwidth}.
As in the simulations in~\cite{leitao2007epidemic}, \msg{IHave} messages are sent immediately without aggregation.
From a leaf source, about 42\% of nodes receive an \msg{IHave} up to about 350\,ms before the \msg{Gossip}.
We therefore set $T_1 = 1000$\,ms, well above this gap so that Plumtree does not graft lazy links during normal operation, and $T_2 = 200$\,ms, the maximum round-trip time.
Plumtree's optimization is disabled by default; Section~\ref{sec:optimize} evaluates the optimization and shorter repair timeouts.
We use Plumtree as the traffic baseline, since it sends no extra payload messages, and \emph{flooding} as the latency baseline.
In flooding, each node forwards the payload to all neighbors except the sender upon first reception, which attains a lower bound on latency without bandwidth limits.
All methods are compared on the same overlay and initial tree.

\subsubsection{Single-broadcast experiments}
In each trial, we (1) generate the overlay with all links eager, (2) have a random root broadcast $m_0$ to form the tree, and (3) have a measurement source broadcast $m_1$; each broadcast runs until no events remain.
For flooding, steps (1)--(2) use Plumtree and only $m_1$ is flooded.
The source of $m_1$ is the \emph{root}, a random \emph{leaf} of the tree (a node with one eager peer), or a \emph{random} node.
We compare Plumtree and flooding with the proposed scheme for $p \in \{0.1, 0.5, 1\}$ and $\theta_h \in \{0, \ldots, 4\}$, and with the decaying variant ($p_h=p\gamma^h$) for $p=1$, $\gamma \in \{0.25, 0.5, 0.75\}$, and the same $\theta_h$, using 50 trials per setting.
Trial $i$ uses seed $i$ and independent random streams for topology generation, source selection, and exuding, so that methods share the overlay, root, and source within a trial.
A uniform random number $u_{m,v,w}$ is fixed for each message $m$, node $v$, and neighbor $w$: node $v$ exudes $m$ to $w$ when $u_{m,v,w}\le p$, or $u_{m,v,w}\le p_h$ for the decaying variant, allowing settings to share the same random draws.
For $m_1$, we measure the mean, median, 95th percentile (p95), and maximum delivery latency over all nodes except the source, the hop count $h_{m_1}(v)$, and the number of messages of each type sent per node.
Since only \msg{Gossip} and \msg{Ex-Gossip} carry the payload, we report both the total number of messages and the number of \emph{payload messages}, which dominate the byte count for large payloads.
Results are averaged over trials; figure error bars show 95\% confidence intervals ($t$-distribution), and differences between settings are reported with paired 95\% confidence intervals, computed from the shared random draws.

\subsubsection{Sequential-broadcast experiments}
The optimization rewires links through payload-less \msg{Graft} messages, so its effect appears only in subsequent broadcasts.
To examine tree changes over successive broadcasts, we also send 20 broadcasts in sequence from random sources after $m_0$.
We compare Plumtree and the proposed scheme ($p=1$, $\theta_h=4$), each with and without the optimization, averaging over the 20 broadcasts and 50 trials.
The optimization threshold is 4, close to the diameter, as recommended for multiple senders~\cite{leitao2007epidemic}.
A node sending an optimization \msg{Graft} message also updates its own peer sets.
We additionally repeat both the single-broadcast and sequential-broadcast experiments with $T_1 \in \{50, 100, 200, 500\}$\,ms, allowing Plumtree to graft lazy links during normal operation (Section~\ref{sec:optimize}).

\subsubsection{Overlay-size and degree experiments}
To examine the effect of overlay topology in single-broadcast trials, we vary the number of nodes as $N \in \{250, 500, 1000, 2000, 4000\}$ with degree $d=8$, and the degree as $d \in \{4, 6, \ldots, 16\}$ with $N=1000$, allowing $\theta_h \le 8$.

\subsubsection{Bandwidth-limited experiments}
In Section~\ref{sec:bandwidth}, we instead model a 100\,Mbit/s uplink at each node, which sends messages one at a time.
Payload messages have size $S+100$ bytes and identifier-only messages have size 100 bytes.
We compare FIFO transmission with priority scheduling for identifier-only messages.
Even in the latter case, an identifier-only message about $m$ is sent after any payload of $m$ that is still queued for the same neighbor.
For the single-broadcast workload, we vary the payload size $S$ from 1\,kB to 1\,MB.
For the concurrent-broadcast workload, random sources issue 100\,kB payloads as a Poisson process at rates of 2 to 20 broadcasts per second.
Each trial issues 100 broadcasts, of which the first 10 are excluded from the statistics, and we run 50 trials.
We use $p=1$, $\theta_h \in \{2, 3, 4\}$, and the default $T_1 = 1000$\,ms.
Apart from the short identifier-only messages, results depend on the bandwidth $B$ only through $S/B$; for example, 100\,kB at 100\,Mbit/s is equivalent to 1\,MB at 1\,Gbit/s.

\subsection{Delivery latency}

\begin{figure*}[!t]
\centering
\includegraphics[width=0.8\textwidth]{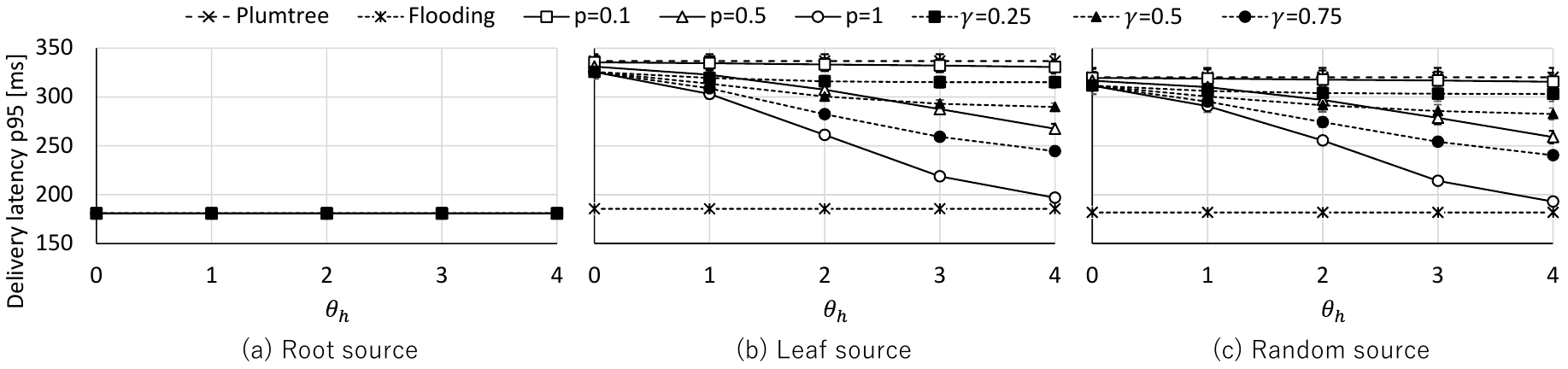}
\caption{p95 latency versus $\theta_h$. Error bars: 95\% confidence intervals.}
\label{fig:latency}
\end{figure*}

\begin{table}[!t]
\centering
\caption{Latency and messages per node. From the root, all settings have the latency shown.}
\label{tab:main}
\footnotesize
\setlength{\tabcolsep}{2.8pt}
\begin{tabular}{lrrrrrrrr}
\hline
Source & $p$ & $\theta_h$ & \multicolumn{4}{c}{Delivery latency [ms]} & \multicolumn{2}{c}{Msgs/node} \\
 & & & mean & median & p95 & max & all & payload \\
\hline
Root & \multicolumn{2}{l}{Plumtree} & 141.3 & 144.8 & 180.8 & 211.1 & 7.00 & 1.00 \\
 & \multicolumn{2}{l}{Flooding} & 141.3 & 144.8 & 180.8 & 211.1 & 7.00 & 7.00 \\
\cline{2-9}
 & 1 & 4 & 141.3 & 144.8 & 180.8 & 211.1 & 9.26 & 3.26 \\
\hline
Leaf & \multicolumn{2}{l}{Plumtree} & 281.0 & 290.8 & 336.8 & 368.5 & 7.00 & 1.00 \\
 & \multicolumn{2}{l}{Flooding} & 145.6 & 148.9 & 185.6 & 216.1 & 7.00 & 7.00 \\
\cline{2-9}
 & 1 & 0 & 246.8 & 256.8 & 325.7 & 364.2 & 7.53 & 1.01 \\
 & 1 & 1 & 211.2 & 216.3 & 303.2 & 352.8 & 7.90 & 1.09 \\
 & 1 & 2 & 176.9 & 175.5 & 261.2 & 325.7 & 8.40 & 1.50 \\
 & 1 & 3 & 157.2 & 157.9 & 218.8 & 285.0 & 9.50 & 2.57 \\
 & 1 & 4 & 149.2 & 151.4 & 196.9 & 245.9 & 10.99 & 4.05 \\
\cline{2-9}
 & 0.5 & 4 & 195.8 & 197.5 & 267.5 & 324.2 & 8.69 & 1.86 \\
 & 0.1 & 4 & 264.3 & 273.8 & 330.8 & 365.7 & 7.32 & 1.04 \\
\hline
Random & \multicolumn{2}{l}{Plumtree} & 264.3 & 274.8 & 320.3 & 352.7 & 7.00 & 1.00 \\
 & \multicolumn{2}{l}{Flooding} & 142.3 & 145.7 & 181.9 & 210.5 & 7.00 & 7.00 \\
\cline{2-9}
 & 1 & 0 & 238.1 & 249.2 & 311.6 & 346.6 & 7.44 & 1.01 \\
 & 1 & 1 & 203.8 & 209.3 & 290.6 & 337.5 & 7.82 & 1.09 \\
 & 1 & 2 & 172.6 & 171.5 & 255.5 & 315.5 & 8.34 & 1.48 \\
 & 1 & 3 & 153.4 & 154.1 & 214.2 & 277.5 & 9.44 & 2.54 \\
 & 1 & 4 & 145.8 & 148.0 & 192.9 & 242.6 & 10.96 & 4.06 \\
\cline{2-9}
 & 0.5 & 4 & 187.8 & 189.5 & 258.9 & 314.0 & 8.72 & 1.91 \\
 & 0.1 & 4 & 250.1 & 259.9 & 315.7 & 350.7 & 7.31 & 1.05 \\
\hline
\end{tabular}

\end{table}

Fig.~\ref{fig:latency} and Table~\ref{tab:main} show the results.
From the root, all methods have exactly the same latency, as the tree is a shortest-delay tree from the root.

From leaf and random sources, Plumtree suffers from the tree detour, with p95 latencies of 336.8 and 320.3\,ms, respectively, whereas flooding achieves 185.6 and 181.9\,ms.

The average p95 latency of exuding decreases monotonically in $p$ and $\theta_h$.
With $p=1$ and $\theta_h=4$, the p95 falls to 196.9\,ms from leaves and 192.9\,ms from random sources, reductions of 41.5\% and 39.8\%.
These values are within 6.1\% of flooding and close about 92\% of the gap between Plumtree and flooding.
With $p=0.1$, the reduction is at most 1.8\%, as exuding then occurs on few links near the source.

\subsection{Message overhead}

\begin{figure*}[!t]
\centering
\includegraphics[width=0.8\textwidth]{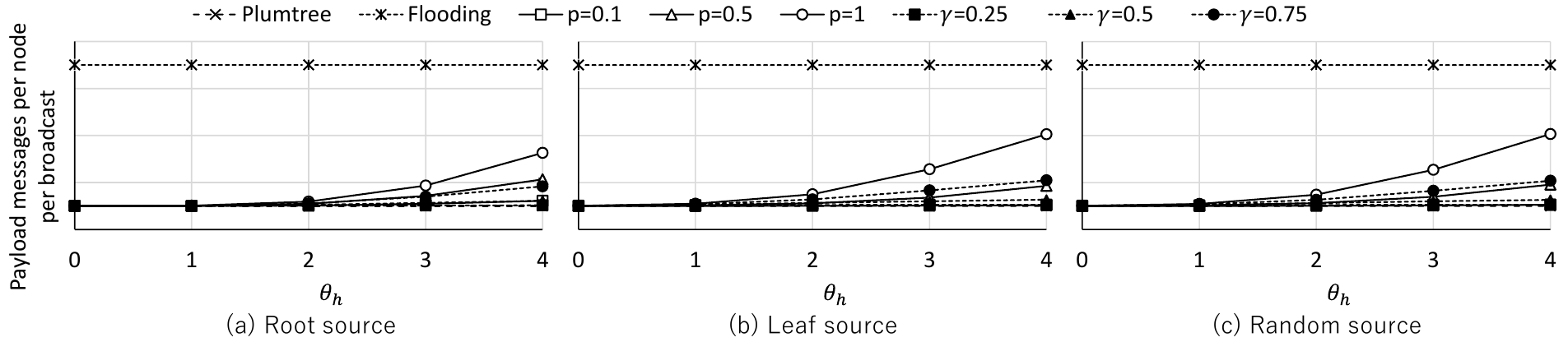}
\caption{Payload messages (\msg{Gossip} and \msg{Ex-Gossip}) per node versus $\theta_h$.}
\label{fig:msg}
\end{figure*}

Plumtree sends 7.00 messages per node: one \msg{Gossip} per tree link and one \msg{IHave} per direction of the 3001 lazy links.
Only 1.00 of them carries the payload, whereas flooding sends 7.00 messages per node, all of which carry the payload.
As Fig.~\ref{fig:msg} shows, the number of payload messages per node under exuding grows with $p$ and $\theta_h$, reaching 4.05 from leaves and 4.06 from random sources with $p=1$ and $\theta_h=4$ (about 58\% of flooding), and stays at or below 1.50 for $\theta_h \le 2$.
For broadcasts from the root, exuding adds 2.26 redundant payload messages per node, bringing the total to 3.26.
Counting all messages, exuding from leaves sends 10.99 per node, more than flooding, as identifier-only messages remain.
In bytes, let $S$ be the payload size and $I$ the size of an identifier-only message, so that a payload message has $S+I$ bytes as in the bandwidth model.
Exuding from leaves then sends $4.05S + 10.99I$ bytes per node, against $7S + 7I$ for flooding and $S + 7I$ for Plumtree.
It thus sends fewer bytes than flooding whenever $S > 1.35I$, and whenever $S > 0.25I$ with $\theta_h = 2$.
The load of the busiest node hardly grows: it sends 8 payload messages at a leaf source, versus 7 at Plumtree's branching nodes.

\begin{figure*}[!t]
\centering
\includegraphics[width=0.8\textwidth]{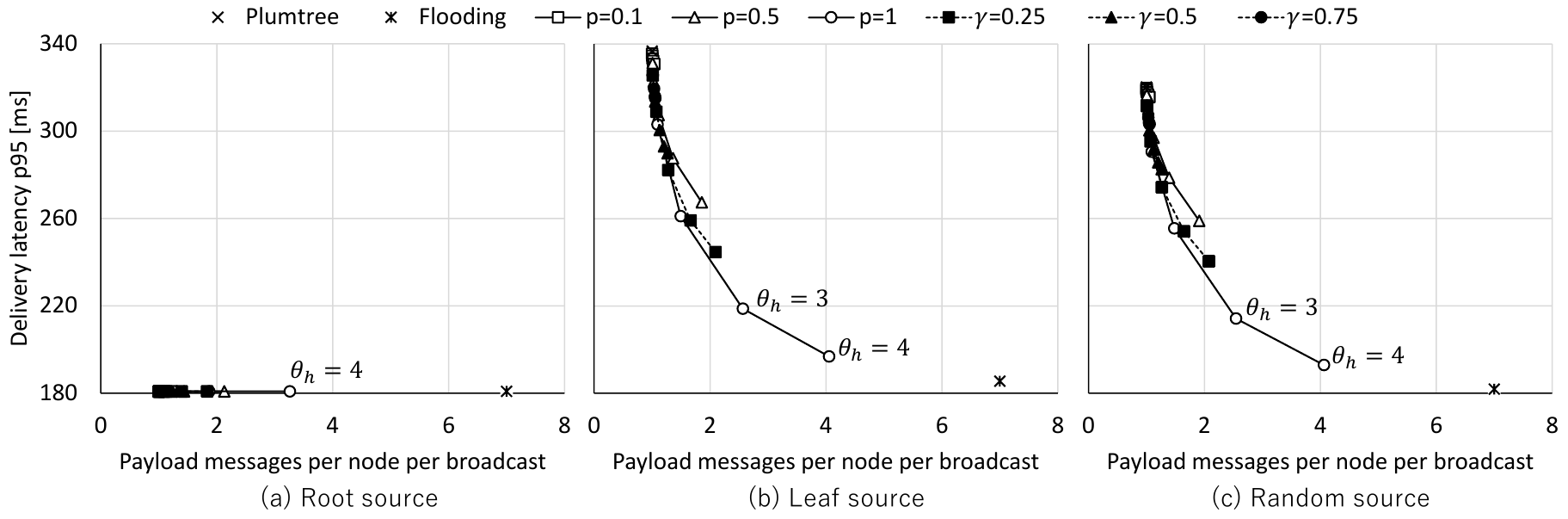}
\caption{p95 latency versus payload messages per node. Points with the same $p$ (solid) or the same $\gamma$ (dashed) are connected in increasing order of $\theta_h$.}
\label{fig:tradeoff}
\end{figure*}

\subsection{Latency--overhead trade-off}

Fig.~\ref{fig:tradeoff} shows that the proposed scheme offers intermediate latency–traffic trade-offs between Plumtree and flooding, with $p$ and $\theta_h$ serving as tuning parameters.
At comparable traffic levels, $p=1$ with a smaller $\theta_h$ tends to outperform a smaller $p$ with a larger $\theta_h$: from leaves, $p=1$ and $\theta_h=2$ yields 261.2\,ms with 1.50 payload messages, versus 267.5\,ms with 1.86 for $p=0.5$ and $\theta_h=4$.
With $p=1$, raising $\theta_h$ from 1 to 2 saves 42.0\,ms for 0.41 more payload messages, whereas raising it from 3 to 4 saves only 21.9\,ms for 1.48 more.

\subsection{Concentrated versus thinly spread exuding}
\label{sec:decay}

Exuding thus works best when concentrated near the source, whose links lead to many nodes still far away along the tree, while farther links increasingly reach nodes already covered.
Of the duplicate payload transmissions (3.05 per node on average) for leaf-source broadcasts with $p=1$ and $\theta_h=4$, 76\% originate from nodes three or four hops from the source, whereas only 6\% originate from nodes within two hops.
The decaying variant tests the opposite strategy: with $\theta_h=4$ and leaf sources, $\gamma=0.25$, 0.5, and 0.75 cut the payload messages to 1.05, 1.27, and 2.09, but the p95 only to 315.2, 289.9, and 244.7\,ms.
When compared at equal payload-traffic levels using linear interpolation, these points lie above the curve connecting the $p=1$ points by 1.1, 5.0, and 7.2\,ms from leaves and by 2.7, 7.2, and 8.1\,ms from random sources.
The differences are significant by bootstrap 95\% confidence intervals, except for $\gamma=0.25$ from leaves.
The points still lie below the curve through the $p=0.5$ points.
With $\theta_h \le 2$, the gaps are within $\pm 4$\,ms.
Hence, for a given budget, exuding with $p=1$ near the source and adjusting $\theta_h$ is preferable to spreading exuding thinly.

\subsection{Tree stability and delivery-time bounds}

In all 4650 trials of Plumtree and the proposed scheme, every node delivered $m_1$, and no \msg{Graft}, \msg{Prune}, or \msg{Gossip-Req} was sent during the measurement broadcast.
Since $m_0$ is flooded without exuding, both methods construct the same spanning tree, which remains unchanged during the measurement broadcast.
A node-by-node check confirmed that the proposition in Section~\ref{sec:bounds} held in all 4,650 trials and in 4,650 additional trials with identical link delays, for which delivery time is proportional to hop count.
In contrast, after the optimization had rewired the tree over ten broadcasts, $t_{\mathrm{tree}}(r,v) > t_{\mathrm{flood}}(r,v)$ held for some nodes in all 50 trials.

\subsection{Comparison with tree rewiring in Plumtree}
\label{sec:optimize}

\begin{table}[!t]
\centering
\caption{Latency and messages per node over 20 broadcasts from random sources, with and without Plumtree's optimization (proposed: $p=1$, $\theta_h=4$). \msg{Prune} counts equal \msg{Graft} counts.}
\label{tab:opt}
\footnotesize
\setlength{\tabcolsep}{3pt}
\begin{tabular}{lrrrrrrr}
\hline
Method & \multicolumn{4}{c}{Delivery latency [ms]} & \multicolumn{3}{c}{Msgs/node} \\
 & mean & median & p95 & max & all & payload & \msg{Graft} \\
\hline
Plumtree & 262.2 & 272.5 & 318.5 & 350.6 & 7.00 & 1.00 & 0.00 \\
Plumtree + opt. & 332.1 & 334.6 & 445.4 & 546.0 & 7.27 & 1.00 & 0.14 \\
Proposed & 145.7 & 147.9 & 193.2 & 242.8 & 10.94 & 4.05 & 0.00 \\
Proposed + opt. & 147.7 & 149.4 & 198.7 & 259.6 & 11.50 & 4.30 & 0.14 \\
\hline
\end{tabular}

\end{table}

Table~\ref{tab:opt} shows the influence of Plumtree's optimization.
Over 20 successive random-source broadcasts, Plumtree's optimization increases p95 latency from 318.5 to 445.4\,ms (+40\%), reaching 485.3\,ms over the final ten broadcasts.
Although it reduces the p95 hop count from 11.7 to 10.6, it rewires about 135 of 999 tree links per broadcast without stabilizing.
We attribute this increase to repeated link replacements that compare hop counts rather than delays and favor the current source. These changes accumulate and disrupt the fastest paths in the tree built by $m_0$.
By contrast, the proposed scheme's p95 changes only from 193.2 to 198.7\,ms (+2.9\%) with optimization.
Identical \msg{Graft} and \msg{Prune} counts confirm that exuding does not affect tree maintenance.

A shorter $T_1$ is another way for Plumtree to use lazy links, but it helps little.
With $T_1 = 50$\,ms, a broadcast from a leaf triggers 0.24 \msg{Graft} messages per node and replaces 46 tree links, yet the p95 drops only from 336.8 to 332.7\,ms.
Presumably, a graft helps little because the payload arrives only after the timeout and a round trip, and because each graft adds only one shortcut link.
Over 20 broadcasts, the changes accumulate, raising the p95 to 359.5 and 367.6\,ms for $T_1 = 50$ and 100\,ms.
The p95 of exuding with $p=1$ and $\theta_h=4$ varies by less than 1\,ms.

\subsection{Overlay size and degree}
\label{sec:scale}

\begin{table}[!t]
\centering
\caption{Overlay size and degree (leaf sources, p95 in ms). $\theta_h$: smallest value ($p=1$) closing 90\% of the gap between Plumtree and flooding; Exuding: P95 with $p=1$ and this $\theta_h$;
Payload: its payload messages per node and share of flooding's extra payload messages.}
\label{tab:scale}
\footnotesize
\setlength{\tabcolsep}{3pt}
\begin{tabular}{rrrrrrrr}
\hline
$N$ & $d$ & Diam. & Plumtree & Flooding & $\theta_h$ & Exuding & Payload \\
\hline
1000 & 4 & 8.6 & 638.5 & 348.9 & 6 & 368.1 & 2.19 (59\%) \\
1000 & 6 & 6.0 & 430.2 & 233.9 & 5 & 243.0 & 3.56 (64\%) \\
250 & 8 & 4.1 & 277.9 & 155.5 & 3 & 165.9 & 4.00 (50\%) \\
500 & 8 & 5.0 & 304.5 & 171.0 & 4 & 177.3 & 4.72 (62\%) \\
1000 & 8 & 5.0 & 336.8 & 185.6 & 4 & 196.9 & 4.05 (51\%) \\
2000 & 8 & 6.0 & 370.9 & 199.6 & 5 & 206.3 & 4.81 (63\%) \\
4000 & 8 & 6.0 & 390.7 & 213.6 & 5 & 224.7 & 4.20 (53\%) \\
1000 & 10 & 5.0 & 292.2 & 160.2 & 4 & 166.7 & 5.60 (57\%) \\
1000 & 12 & 4.0 & 253.9 & 140.1 & 4 & 144.6 & 7.07 (61\%) \\
1000 & 14 & 4.0 & 236.1 & 128.1 & 4 & 131.2 & 8.65 (64\%) \\
1000 & 16 & 4.0 & 218.7 & 118.8 & 4 & 121.5 & 10.30 (66\%) \\
\hline
\end{tabular}

\end{table}

Table~\ref{tab:scale} shows the results.
From 250 to 4000 nodes and degrees 4 to 16, the p95 of Plumtree is 1.78 to 1.86 times that of flooding from leaves, and 1.68 to 1.81 times from random sources, so the detour does not shrink with scale.
The smallest $\theta_h$ closing 90\% of the gap grows with the diameter and is in most cases one less than the diameter; it equals the diameter for $d \ge 12$ and is 6 for the diameter of 8.6 with $d=4$.
This $\theta_h$ requires 50 to 66\% of flooding's extra payload messages in every overlay.
A fixed $\theta_h=2$ closes less of the gap as the diameter grows: 72\%, 50\%, and 39\% for $N=250$, 1000, and 4000.

\subsection{Bandwidth limits and concurrent broadcasts}
\label{sec:bandwidth}

\begin{figure}[!t]
\centering
\includegraphics[width=0.8\columnwidth]{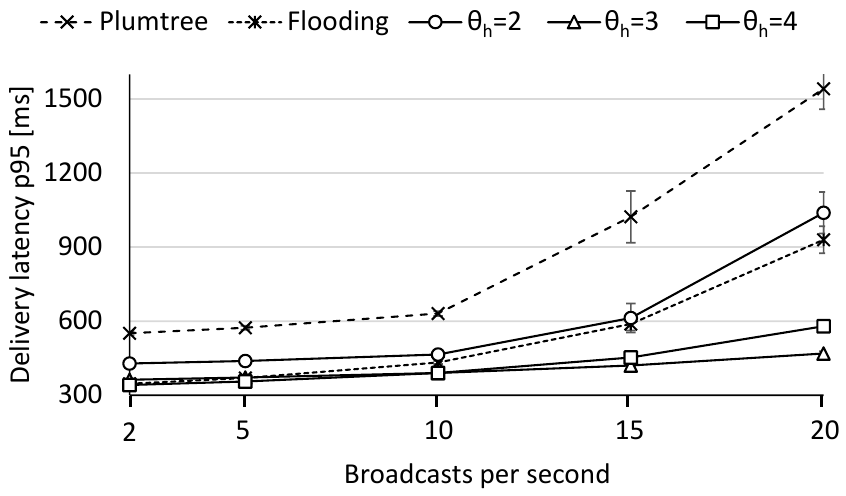}
\caption{p95 latency of concurrent broadcasts versus their rate (100\,kB payloads, 100\,Mbit/s uplinks, priority for identifier-only messages; exuding with $p=1$). Error bars: 95\% confidence intervals.}
\label{fig:concurrent}
\end{figure}

For single broadcasts from leaf sources with FIFO uplinks, the p95 latencies for Plumtree, exuding ($p=1$, $\theta_h=4$), and flooding are 561.7, 336.3, and 337.0\,ms, respectively, for 100\,kB payloads, and 2176, 1417, and 1309\,ms for 1\,MB payloads.
The corresponding p95 latencies with priority scheduling differ from those with FIFO by at most 6\,ms.

Under concurrent broadcasts (Fig.~\ref{fig:concurrent}), Plumtree degrades first.
At 15 broadcasts per second, its payload messages occupy only 13\% of the uplink capacity on average, yet its p95 rises to 1023\,ms.
Every broadcast crosses the same branch nodes: the busiest 1\% of the nodes send 7.00 payload messages per broadcast, as many as every node does in flooding.
The resulting queueing delays cause $T_1$ to expire, and after 4138 grafts, the eager-link graph contains an average of 192 connected components rather than a single spanning tree, although repair still delivers every message.
Exuding degrades more gracefully, presumably because it uses the fastest of many paths.
With $\theta_h=3$, the p95 is 363, 391, 420, and 469\,ms at 2, 10, 15, and 20 broadcasts per second, and stays below that of flooding from 10 broadcasts per second onward.
Flooding exceeds the uplink capacity at 20 per second.
The best $\theta_h$ decreases with the load: 4 up to 5 broadcasts per second, 3 and 4 tied at 10, and 3 from 15 onward.
With $\theta_h=2$, fewer nodes receive exuded copies, so once repairs start, its latency follows the tree delays.
Repair timers also expire with exuding under such load, but less often: at 15 per second, $\theta_h=3$ sends 2011 \msg{Graft} messages and leaves 90 components.
With FIFO uplinks, identifier-only messages wait behind exuded payloads: at 15 per second, $\theta_h=4$ sends 8930 \msg{Graft} messages and leaves 285 components, versus 2184 and 100 with priority, so identifier-only messages should be prioritized.

\section{Discussion and Limitations}
\label{sec:discussion}

\textbf{Per-source trees.}
Under the idealized conditions of Section~\ref{sec:bounds}, a tree per source achieves the same delivery latency as flooding.
The first broadcast from each source sends 7 payload messages per node, while subsequent broadcasts from the same source send only 1.
In terms of cumulative payload messages per source, per-source trees become more efficient than exuding from the second broadcast for $\theta_h=4$ and from the thirteenth for $\theta_h=2$.
However, per-source trees require peer sets and repairs per source, increasing state and signaling.
Their initial flooding cost is poorly amortized when many peers originate few broadcasts, as may occur in blockchain transaction dissemination; exuding instead retains a single shared tree.

\textbf{Choice of parameters.}
With $p=1$, $\theta_h$ should be selected from the trade-off curve according to the available traffic budget.
A value about one below the diameter closes 90\% of the gap, and $\theta_h=2$ already cuts the p95 by about 20\% with 1.5 payload messages per node for $N=1000$ and $d=8$.
The value can be set from the hop counts of past broadcasts and lowered under heavy load.
Exuding is not switched off at the root, which may end up near a leaf as nodes join and leave.
Identifier-only messages should be sent ahead of payloads to avoid spurious repairs.

\textbf{Scope.}
We simulated static random regular overlays with independent uniform delays and uplink-only bandwidth limits, without failures, churn, or congestion inside the network.
In latency-aware overlays, e.g., those built by X-BOT~\cite{xbot2009}, tree detours, and thus the gains, may be smaller.
In the presence of message loss, tree messages are repaired as in Plumtree; when a node receives \msg{Mid-Gossip} before obtaining the corresponding payload, a \msg{Gossip-Req}/\msg{Gossip} exchange suffices if both messages are delivered.
Whether exuding also speeds up recovery, as in PRM~\cite{banerjee2006resilient}, remains to be measured.

\section{Conclusion}

Probabilistic exuding accelerates multi-source broadcasts in Plumtree through off-tree forwarding without interfering with tree maintenance.
In simulations, it reduced p95 latency from leaf and random sources by about 40\%, approaching flooding latency with only 58\% of its payload messages.
The benefits persisted across overlay sizes, degrees, and bandwidth-limited workloads.
Future work includes evaluating the method in the presence of failures and churn and investigating adaptive selection of $\theta_h$.

\section*{Use of AI Disclosure}
Anthropic Claude Opus 5.5 and OpenAI GPT-6 were used to assist with simulator implementation, debugging, and manuscript proofreading.
All AI-assisted code and text revisions were reviewed by the author.

\bibliographystyle{IEEEtran}
\bibliography{IEEEabrv,reference}

@STRING{IEEE_J_JSAC       = "{IEEE} J. Sel. Areas Commun."}

@STRING{IEEE_J_MC         = "{IEEE} Trans. Mobile Comput."}

@STRING{IEEE_J_NET        = "{IEEE/ACM} Trans. Netw."}

@STRING{IEEE_O_CSTO        = "{IEEE} Commun. Surveys Tuts."}

@inproceedings{demers1987epidemic,
  author    = {Alan Demers and Dan Greene and Carl Hauser and Wes Irish and John Larson and Scott Shenker and Howard Sturgis and Dan Swinehart and Doug Terry},
  title     = {Epidemic Algorithms for Replicated Database Maintenance},
  booktitle = {ACM PODC},
  pages     = {1--12},
  year      = {1987}
}

@article{birman1999bimodal,
  author  = {Kenneth P. Birman and Mark Hayden and Oznur Ozkasap and Zhen Xiao and Mihai Budiu and Yaron Minsky},
  title   = {Bimodal Multicast},
  journal = {ACM Trans. Comput. Syst.},
  volume  = {17},
  number  = {2},
  pages   = {41--88},
  year    = {1999}
}

@inproceedings{jannotti2000overcast,
  author    = {John Jannotti and David K. Gifford and Kirk L. Johnson and M. Frans Kaashoek and James W. {O'Toole, Jr.}},
  title     = {Overcast: Reliable Multicasting with an Overlay Network},
  booktitle = {USENIX OSDI},
  pages     = {197--212},
  year      = {2000}
}

@article{chu2002case,
  author  = {{Yang-hua} Chu and Sanjay G. Rao and Srinivasan Seshan and Hui Zhang},
  title   = {A Case for End System Multicast},
  journal = IEEE_J_JSAC,
  volume  = {20},
  number  = {8},
  pages   = {1456--1471},
  year    = {2002}
}

@inproceedings{banerjee2002scalable,
  author    = {Suman Banerjee and Bobby Bhattacharjee and Christopher Kommareddy},
  title     = {Scalable Application Layer Multicast},
  booktitle = {ACM SIGCOMM},
  pages     = {205--217},
  year      = {2002}
}

@article{castro2002scribe,
  author  = {Miguel Castro and Peter Druschel and Anne-Marie Kermarrec and Antony Rowstron},
  title   = {{SCRIBE}: A Large-Scale and Decentralized Application-Level Multicast Infrastructure},
  journal = IEEE_J_JSAC,
  volume  = {20},
  number  = {8},
  pages   = {1489--1499},
  year    = {2002}
}

@article{eugster2003lightweight,
  author  = {Patrick Th. Eugster and Rachid Guerraoui and Sidath B. Handurukande and Petr Kouznetsov and Anne-Marie Kermarrec},
  title   = {Lightweight Probabilistic Broadcast},
  journal = {ACM Trans. Comput. Syst.},
  volume  = {21},
  number  = {4},
  pages   = {341--374},
  year    = {2003}
}

@inproceedings{rodrigues2003adaptive,
  author    = {Lu{\'\i}s Rodrigues and Sidath Handurukande and Jos{\'e} Pereira and Rachid Guerraoui and Anne-Marie Kermarrec},
  title     = {Adaptive Gossip-Based Broadcast},
  booktitle = {DSN},
  pages     = {47--56},
  year      = {2003}
}

@inproceedings{tang2005gocast,
  author    = {Chunqiang Tang and Rong N. Chang and Christopher Ward},
  title     = {{GoCast}: Gossip-Enhanced Overlay Multicast for Fast and Dependable Group Communication},
  booktitle = {DSN},
  pages     = {140--149},
  year      = {2005}
}

@article{banerjee2006resilient,
  author  = {Suman Banerjee and Seungjoon Lee and Bobby Bhattacharjee and Aravind Srinivasan},
  title   = {Resilient Multicast Using Overlays},
  journal = IEEE_J_NET,
  volume  = {14},
  number  = {2},
  pages   = {237--248},
  year    = {2006}
}

@inproceedings{leitao2007hyparview,
  author    = {Jo{\~a}o Leit{\~a}o and Jos{\'e} Pereira and Lu{\'\i}s Rodrigues},
  title     = {{HyParView}: A Membership Protocol for Reliable Gossip-Based Broadcast},
  booktitle = {IEEE/IFIP DSN},
  pages     = {419--429},
  year      = {2007}
}

@inproceedings{leitao2007epidemic,
  author    = {Jo{\~a}o Leit{\~a}o and Jos{\'e} Pereira and Lu{\'\i}s Rodrigues},
  title     = {Epidemic Broadcast Trees},
  booktitle = {IEEE SRDS},
  pages     = {301--310},
  year      = {2007}
}

@article{hosseini2007survey,
  author  = {Mojtaba Hosseini and Dewan Tanvir Ahmed and Shervin Shirmohammadi and Nicolas D. Georganas},
  title   = {A Survey of Application-Layer Multicast Protocols},
  journal = IEEE_O_CSTO,
  volume  = {9},
  number  = {3},
  pages   = {58--74},
  year    = {2007}
}

@article{li2009fast,
  author  = {Zhenyu Li and Zengyang Zhu and Gaogang Xie and Zhongcheng Li},
  title   = {Fast and Proximity-Aware Multi-Source Overlay Multicast under Heterogeneous Environment},
  journal = {Comput. Commun.},
  volume  = {32},
  number  = {2},
  pages   = {257--267},
  year    = {2009}
}

@misc{vyzovitis2020gossipsub,
  author       = {Dimitris Vyzovitis and Yusef Napora and Dirk McCormick and David Dias and Yiannis Psaras},
  title        = {{GossipSub}: Attack-Resilient Message Propagation in the {Filecoin} and {ETH2.0} Networks},
  howpublished = {arXiv preprint arXiv:2007.02754},
  year         = {2020}
}

@inproceedings{scheidt2023pubsub,
  author    = {Flaviene {Scheidt de Cristo} and Wazen M. Shbair and Lucian Trestioreanu and Radu State},
  title     = {Pub/Sub Dissemination on the {XRP} Ledger},
  booktitle = {IEEE LATINCOM},
  pages     = {1--6},
  year      = {2023}
}

@article{ren2024intelligent,
  author  = {Zhe Ren and Xinghua Li and Yinbin Miao and Zhuowen Li and Zihao Wang and Mengyao Zhu and Ximeng Liu and Robert H. Deng},
  title   = {Intelligent Adaptive Gossip-Based Broadcast Protocol for {UAV-MEC} Using Multi-Agent Deep Reinforcement Learning},
  journal = IEEE_J_MC,
  volume  = {23},
  number  = {6},
  pages   = {6563--6578},
  year    = {2024}
}

@inproceedings{wooders2024cloudcast,
  author    = {Sarah Wooders and Shu Liu and Paras Jain and Xiangxi Mo and Joseph E. Gonzalez and Vincent Liu and Ion Stoica},
  title     = {Cloudcast: High-Throughput, Cost-Aware Overlay Multicast in the Cloud},
  booktitle = {USENIX NSDI},
  pages     = {281--296},
  year      = {2024}
}

@inproceedings{scheidt2024nine,
  author    = {Flaviene {Scheidt de Cristo} and Jorge Augusto Meira and Jean-Philippe Eisenbarth and Radu State},
  title     = {A 9-Dimensional Analysis of {GossipSub} over the {XRP} Ledger Consensus Protocol},
  booktitle = {IEEE/IFIP NOMS},
  pages     = {1--7},
  year      = {2024},
  doi       = {10.1109/NOMS59830.2024.10575688}
}

@inproceedings{xbot2009,
  author    = {Jo{\~a}o Leit{\~a}o and Jo{\~a}o Pedro Marques and Jos{\'e} Pereira and Lu{\'\i}s Rodrigues},
  title     = {{X-BOT}: A Protocol for Resilient Optimization of Unstructured Overlays},
  booktitle = {IEEE SRDS},
  pages     = {236--245},
  year      = {2009},
  doi       = {10.1109/SRDS.2009.20}
}

@inproceedings{thicket2010,
  author    = {M{\'a}rio Ferreira and Jo{\~a}o Leit{\~a}o and Lu{\'\i}s Rodrigues},
  title     = {Thicket: A Protocol for Building and Maintaining Multiple Trees in a {P2P} Overlay},
  booktitle = {IEEE SRDS},
  pages     = {293--302},
  year      = {2010},
  doi       = {10.1109/SRDS.2010.19}
}

@inproceedings{brisa2012,
  author    = {Miguel Matos and Valerio Schiavoni and Pascal Felber and Rui Oliveira and Etienne Rivi{\`e}re},
  title     = {{BRISA}: Combining Efficiency and Reliability in Epidemic Data Dissemination},
  booktitle = {IEEE IPDPS},
  pages     = {983--994},
  year      = {2012},
  doi       = {10.1109/IPDPS.2012.92}
}

\end{document}